\documentclass{article}
\usepackage[margin=1in]{geometry}
\usepackage[hidelinks]{hyperref}
\usepackage{parskip}
\usepackage{amsmath}
\usepackage{amssymb}
\usepackage{natbib}
\usepackage{float}
\usepackage{graphicx}
\usepackage{tikz}
\tikzset{ntp/.style={circle, thin, minimum size=2mm, inner sep=0, fill=white, #1}}
\usepackage{color}
\title{On possible issues of Backus average}
\author{
Ayiaz Kaderali\footnote{%
Department of Earth Sciences, Memorial University of Newfoundland, Canada, {\tt ayiazkaderali@gmail.com}}\,,
Izabela Kudela\footnote{%
Department of Earth Sciences, Memorial University of Newfoundland, Canada,
{\tt izabelakudela@gmail.com}}\,,  
Theodore Stanoev\footnote{%
Department of Earth Sciences, Memorial University of Newfoundland, Canada,
{\tt theodore.stanoev@gmail.com}}
}
\date{}
\begin{document}
\maketitle
\begin{abstract}
In this paper, we continue the study of~\citet{BosEtAl2018} regarding statistical and numerical considerations of the~\citet{Backus1962} product approximation.
While the approximation is typically quite good for seismological scenarios,~\citet{BosEtAl2018} demonstrate a physical scenario that could, in spite of the stability conditions for isotropic media, lead to an issue within the Backus average.
Using the Preliminary Reference Earth Model of~\citet{DziewonskiAnderson1981} and a case study in the upper oceanic crust, we investigate whether this issue is likely to occur in the context of seismology.
\end{abstract}
\section{Introduction}
The Backus average is a method that produces a homogenous medium that is long-wave equivalent to an inhomogeneous stack of thin layers.
Notwithstanding the ubiquitous acceptance of the Backus average, it has been the topic of recent study for~\citet{AdamusEtAl2018} and~\citet{BosEtAl2017,BosEtAl2018,BosEtAl2019} as well as~\cite{DaltonEtAl2019} and~\citet{DaltonKaderali2020}.
While the mathematical underpinnings of the Backus approach are analyzed by~\citet{BosEtAl2017}, there may exist a possible issue with the sole mathematical approximation used by Backus~\citep{BosEtAl2018}.


In spite of stability conditions,~\citet{BosEtAl2018} demonstrate that it is mathematically and physically possible for the relative error of the Backus product approximation to equal 100\%\,.
Herein, we show that it is unlikely for such an error to occur within the context of seismology.
\section{Product approximation}
\label{sec:ProdApprox}
Let us consider the product approximation of~\citet{Backus1962}, which states that
\begin{quote}
[t]he only approximation that [he makes] in the present paper is the following: if~$f(x_{3})$ is nearly constant when~$x_{3}$ changes by no more than~$\ell'$\,, while~$g(x_{3})$ may vary by a large fraction of this distance, then, approximately,
\begin{equation}\label{eq:BackusApprox}
\overline{f\,g}\approx\overline{f}\,\,\overline{g}
\,.
\end{equation}
\end{quote}

Using the formulation of~\citet{BosEtAl2018}, which states that
\begin{quote}
the difference between the average of the product and the product of the averages is
\begin{equation}\label{eq:BAappDiff}
E\left(f\,,\,g\right) :=
\overline{f\,g} -
\overline{f}\,\,\overline{g}
\,,
\end{equation}
where, for any vector~$\overline{x}\in\mathbb{R}^{n}$\,, [they] set
\begin{equation*}
\overline{x} :=
\sum\limits_{k=1}^{n}w_{k}\,x_{k}
\,.
\end{equation*}
\end{quote}

The relative error is
\begin{equation}\label{eq:BAppRelError}
R\left(f\,,\,g\right) =
\frac{E\left(f\,,\,g\right)}{\overline{f\,g}}\times100\%\,.
\end{equation}
It follows that if~$\overline{g}=0$ then~$R\left(f\,,\,g\right)=100\%$\,.
To examine the consequences of~$\overline{g}=0$\,, in the context of layers composed of isotropic Hookean solids, expressions for~$f$ and~$g$ may be obtained from the isotropic stress-strain relations~\citep[Section~3.6]{BosEtAl2018}.
We find that~$f$ corresponds to lateral-strain-tensor components that are assumed to be nearly constant, whereas
\begin{equation}\label{eq:gIso}
g = 
\frac{c_{1111}-2\,c_{2323}}{c_{1111}}
\end{equation}
corresponds to elasticity parameters that rapidly vary from layer to layer.

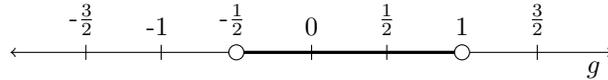
\begin{figure}[h]
\centering
\begin{tikzpicture}
%
%
\draw[<->] (-4,0)--(4,0);
\node[below] at (3.75,0) {$g$};
%
%
\draw (-3,-0.1)--(-3,0.1);
\draw (-2,-0.1)--(-2,0.1);
\draw (-1,-0.1)--(-1,0.1);
\draw (0,-0.1)--(0,0.1);
\draw (1,-0.1)--(1,0.1);
\draw (2,-0.1)--(2,0.1);
\draw (3,-0.1)--(3,0.1);
%
%
\node[above] at (-3.05,0.1) {-$\tfrac{3}{2}$};
\node[above] at (-2.05,0.1) {-1};
\node[above] at (-1.05,0.1) {-$\tfrac{1}{2}$};
\node[above] at (0,0.1) {0};
\node[above] at (1,0.1) {$\tfrac{1}{2}$};
\node[above] at (2,0.1) {1};
\node[above] at (3,0.1) {$\tfrac{3}{2}$};
%
%
\draw[very thick] (-1,0) node[ntp={draw}] {} -- (2,0) node[ntp={draw}] {};
\end{tikzpicture}
\caption{
The value of~$g$ approaches a maximum of 1 when~$c_{2323}$ is at a minimum.
Conversely, the value of~$g$ approaches a minimum of~$-\tfrac{1}{2}$ when~$c_{2323}$ is at a maximum; thus,~$g\in\left(-\tfrac{1}{2}\,,\,1\right)$\,.
}
\label{fig:gRange}
\end{figure}

Let us examine the stability conditions for isotropic media to determine the range of physically possible values of~$g$\,. 
Herein, the stability conditions may derived from the stress-strain relations for isotropy, which are
\begin{equation}\label{eq:StressStrainIso}
\sigma_{ij}
=
\lambda\,\delta_{ij}\sum\limits^{3}_{k=1}\varepsilon_{kk} +
2\,\mu\,\varepsilon_{ij}
\,,\quad
i,j\in\left\{1,2,3\right\}
\,,
\end{equation}
where~$\lambda$ and~$\mu$ are the Lam{\'e} parameters and are defined as
$
\lambda:=c_{1111}-2\,c_{2323}
\,\,{\rm and}\,\,
\mu:=c_{2323}
$\,. 
As it is a requirement for all symmetric positive-definite matrices, all of its eigenvalues must be positive~\citep[see e.g.][Theorem~4.3.2]{Slawinski2020a}.
Thus, for isotropy, the stability conditions are
\begin{equation}\label{eq:StabCondIso}
c_{1111}>0\,,\quad
c_{2323}>0\,,\quad
c_{1111}>\tfrac{4}{3}c_{2323}
\,.
\end{equation}
By applying expressions~(\ref{eq:StabCondIso}) to expression~(\ref{eq:gIso}), we deduce that~$g$ is positive when~$c_{1111}>2\,c_{2323}$ and that~$g$ is negative when~$\frac{4}{3}c_{2323}<c_{1111}<2\,c_{2323}$\,; the range of~$g$ is illustrated in Figure~\ref{fig:gRange}.

Since~$g$ can be either negative or positive, and the elasticity parameters---by the stability conditions---are continuous and positive, we conclude that it is possible for~$g$ to equal zero.


Considering~\citet[Exercise~5.13]{Slawinski2020a}, we might obtain Poisson's ratio in terms of the Lam\'e parameters,
\begin{equation}\label{eq:PoissonLame}
\nu
=
\frac{\lambda}{2\left(\lambda+\mu\right)}
\,,
\end{equation}
which is the desired expression.
Alternatively, we might obtain expression~\eqref{eq:PoissonLame} by using the relations among Poisson's ratio, Young's modulus and the Lam\'e parameters~\citep[see e.g.][Remark~5.14.7]{Slawinski2020a}.

For a two-dimensional case, expression~\eqref{eq:PoissonLame} becomes\footnote{%
	We obtain the two-dimensional Poisson's ratio from the transverse stress component of Hooke's law, expressed in the $x_1x_3$-plane\,, $\sigma_{11} = 0$\,, and solving for $\nu=-\varepsilon_{11}/\varepsilon_{33}$\,.
}
\begin{equation}\label{eq:PoissonLame2D}
\nu
=
\frac{\lambda}{\lambda+2\,\mu}
\,,
\end{equation}
which is equivalent to expression~(\ref{eq:gIso}).
Notably, the properties of a transversely isotropic medium are captured by two-dimensional model that contains the rotation-symmetry axis.
Expression~\eqref{eq:PoissonLame2D} might be useful considering the fact that the Backus average produces a homogeneous transversely isotropic medium that is long-wave equivalent to a stack of thin isotropic layers.

The range of possible values of~$\nu$ in expression~(\ref{eq:PoissonLame}) are determined by the stability conditions, which are determined from the eigenvalues of the positive-definite elasticity tensor used therein.
Thus, the stability conditions for isotropy, in terms of~$\lambda$ and~$\mu$\,, are
\begin{equation}\label{eq:StabCondLame}
\lambda > -\frac{2}{3}\mu
\quad{\rm and}\quad
\mu > 0
\,.
\end{equation}

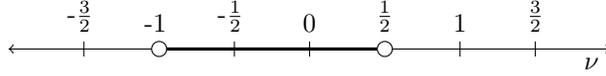
\begin{figure}[h]
\centering
\begin{tikzpicture}
%
%
\draw[<->] (-4,0)--(4,0);
\node[below] at (3.75,0) {$\nu$};
%
%
\draw (-3,-0.1)--(-3,0.1);
\draw (-2,-0.1)--(-2,0.1);
\draw (-1,-0.1)--(-1,0.1);
\draw (0,-0.1)--(0,0.1);
\draw (1,-0.1)--(1,0.1);
\draw (2,-0.1)--(2,0.1);
\draw (3,-0.1)--(3,0.1);
%
%
\node[above] at (-3.05,0.1) {-$\tfrac{3}{2}$};
\node[above] at (-2.05,0.1) {-1};
\node[above] at (-1.05,0.1) {-$\tfrac{1}{2}$};
\node[above] at (0,0.1) {0};
\node[above] at (1,0.1) {$\tfrac{1}{2}$};
\node[above] at (2,0.1) {1};
\node[above] at (3,0.1) {$\tfrac{3}{2}$};
%
%
\draw[very thick] (-2,0) node[ntp={draw}] {} -- (1,0) node[ntp={draw}] {};
\end{tikzpicture}
\caption{
The value of~$\nu$ approaches a maximum of~$\tfrac{1}{2}$ when~$\mu$ is at a minimum.
Conversely, the value of~$\nu$ approaches a minimum of~$-1$ when~$\mu$ is at a maximum; thus,~$\nu\in\left(-1\,,\,\tfrac{1}{2}\right)$\,.
}
\label{fig:nuRange}
\end{figure}

Considering the ranges of values of expressions~\eqref{eq:gIso} and~\eqref{eq:PoissonLame}, we reckon that if (a)~$g>0$ then $\nu>0$\,, (b)~$g=0$ then $\nu=0$\,, and (c)~$g<0$ then $\nu<0$\,.
For naturally occurring solids, $\nu>0$\,; the ratio being positive means that the diminishing of a cylinder's length is being accompanied by the extension of its radius~\citep[e.g.][p.~204]{Slawinski2020a}.
Hence, the range illustrated in Figure~\ref{fig:nuRange} reduces to~$\nu\in\left(0\,,\tfrac{1}{2}\right)$\,.
\section{Seismological examples}
To gain insight into whether or not we might encounter $\overline{g}=0$\,, let us consider two seismological examples.

The Preliminary Reference Earth Model (PREM) of~\citet{DziewonskiAnderson1981} is a one-dimensional model that presents the properties of the Earth as a function of depth.
The PREM is a mathematical analogy that serves as a background model for the planet as a whole; it assumes spherical symmetry in order to subdivide the interior of the Earth into nine principal regions.
This model establishes Earth-specific properties that include  density, $\rho$\,, and $P$- and $S$-wave speeds, which are 
\begin{equation}\label{eq:PSwaveSpeeds}
v_{P} 
=
\sqrt{\frac{\lambda+2\,\mu}{\rho}}
\quad{\rm and}\quad
v_{S} 
=
\sqrt{\frac{\mu}{\rho}}
\,.
\end{equation}
These nine principal regions are distinguished from one another by a rapid change in speeds of $P$ and $S$ waves along interfaces, which indicates a diverse range of elastic properties within the medium.
The speeds of the irrotational and equivoluminal waves are functions of the different elasticity parameters and, hence, propagate at different speeds within the model.

\begin{figure}[h]
\centering
\includegraphics[width=0.8\textwidth]{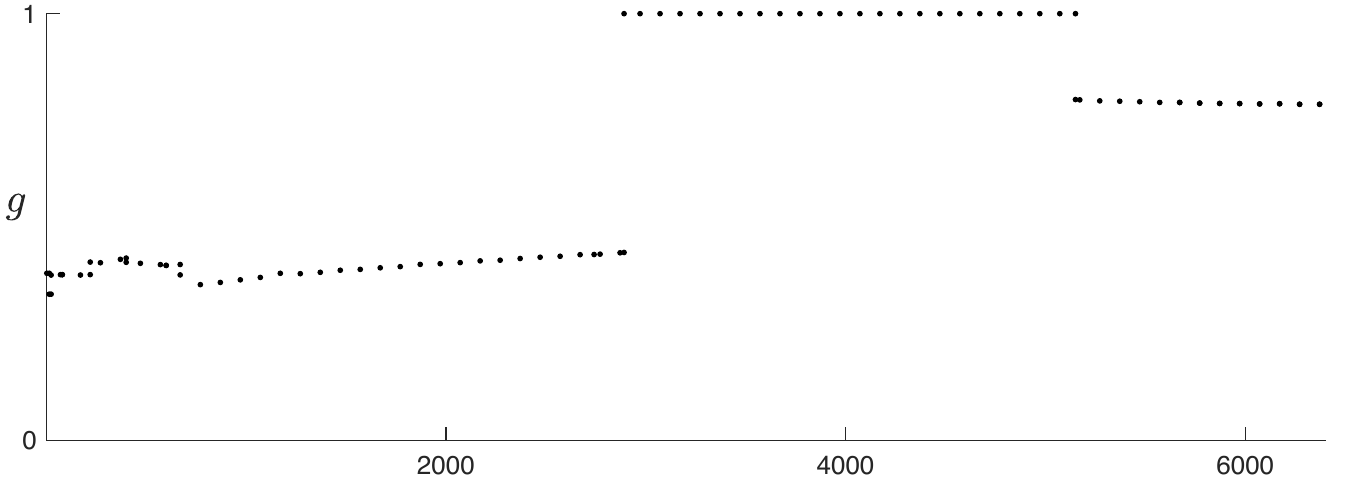}
\caption{$g$ as a function of depth (km) using PREM}
\label{fig:g}
\end{figure}

Let us consider data from~\citet[Table~1]{Bormann2012}, which lists~84 samples of---among other parameters---$v_{P}$\,, $v_{S}$\,, and $\rho$ as functions of depth ranging from 0 to 6371\,km for an isotropic PREM.
In view of the relationship between Lam\'e and elasticity parameters, we may compute
\begin{equation}\label{eq:c1111c2323}
c_{1111} = c^{\ast}_{1111}/\rho = v_{P}^{2}
\quad{\rm and}\quad
c_{2323} = c^{\ast}_{2323}/\rho = v_{S}^{2}
\end{equation}
for each of the 84 samples.
Herein, $c_{1111}$ and $c_{2323}$ are the elasticity parameters scaled by density, $\rho$\,, as opposed to their non-scaled counterparts, $c^{\ast}_{1111}$ and $c^{\ast}_{2323}$\,.
We use the $P$- and $S$-wave speeds, along with density, to calculate the density-scaled elasticity parameters to plot the values of $g$\,, in expression~\eqref{eq:gIso}, as a function of depth in Figure~\ref{fig:g}.
Therein, the resultant points of discontinuity arise from the rapid change in speed of $P$ and $S$ waves across the interfaces of the principal regions.

For samples between 2891\,km and 5150\,km,~$v_{S}=0$\,.
Since the propagation of~$S$ waves requires a material to have rigidity, and liquids are absent of rigidity, we interpret this range of samples to correspond to the outer core.
Since the~$S$-wave speed equals zero,~$c_{2323}=0$\,; consequently, $g=1$\,.

From Figure~\ref{fig:g}, we observe that $g>0$ throughout and, thus, deduce that $\overline{g}$ cannot equal zero.
Therefore, following the conclusions of Section~\ref{sec:ProdApprox}, our results support that $g>0$ for naturally occurring solids within an isotropic PREM.
Hence, we may conclude that it is improbable for the relative error of the Backus average approximation to equal 100\% for such a model.
With that being said, the average Earth model is limited, especially in the upper regions of the Earth.
\citet[Section 2]{DziewonskiAnderson1981} indicate the lateral heterogeneity in the first few tens of kilometres is so large that an average model does not reflect the actual Earth structure at any point.
Thus, to assess the likelihood of encountering $\overline{g} = 0$ in shallow regions of the Earth, i.e., upper oceanic crust, let us turn our attention to the second seismological example.
 
\begin{figure}[H]
\centering
\includegraphics[width=0.8\textwidth]{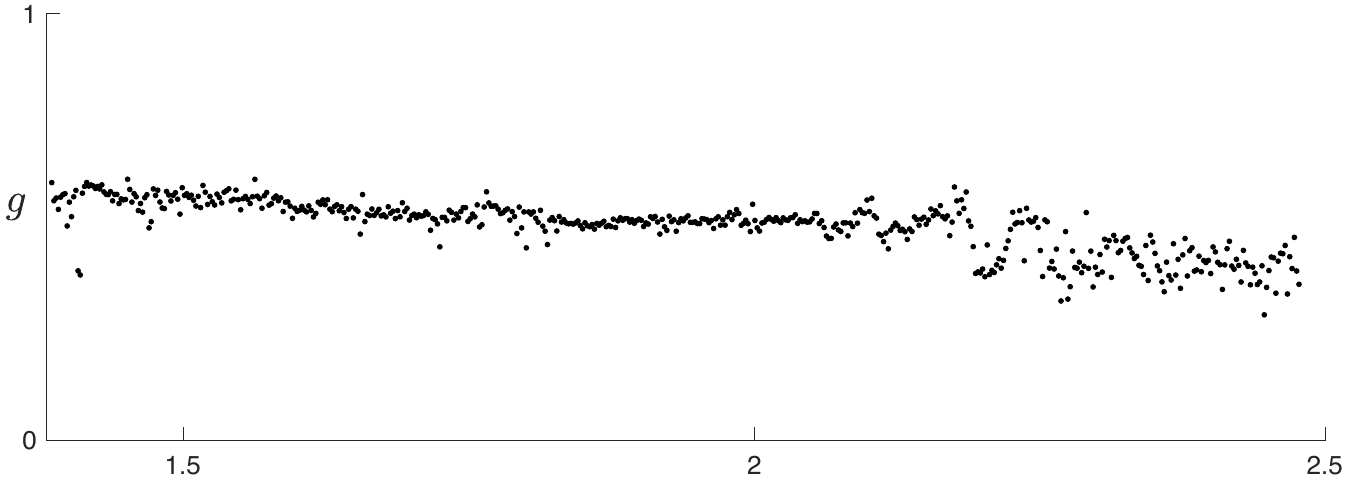}
\caption{$g$ as a function of depth (km) using well log data}
\label{fig:g_ZhouKaderali2006}
\end{figure}

We consider the case study of~\citet{ZhouKaderali2006}, which describes the acquisition of walkaway VSP data in a region offshore Eastern Canada.
The region is horizontally layered, which is required for~\citeauthor{Backus1962} averaging, and is comprised mostly of shales.
The $P$- and $S$-wave speeds are obtained from compressional- and shear-sonic logs to a depth of approximately two and a half kilometres.
We use the density-scaled elasticity parameters for the calculation of $g$\,, since the reliability of density logs is questionable.
In Figure~\ref{fig:g_ZhouKaderali2006}, we plot the values of $g$ for increasing depths and observe that $g>0$ for all depths, which indicates that $\overline{g}$ cannot equal zero.
Notably, the mean value of $g$ is $\overline{g} = 0.5033 \pm 0.0598$\,, and the mean Poisson's ratio, using expression~\eqref{eq:PoissonLame}, is $\overline{\nu} = 0.3337 \pm 0.0276$\,.
The small standard deviations of these quantities indicate that the considered subsurface is, indeed, comprised mostly of similar material.
The increasing scatter in the $P$ and $S$ waves observed with depth, in Figure~\ref{fig:g_ZhouKaderali2006}, may be due to the fine sampling rate of the sonic logs.
Since $g$ and $\nu$ are functions of the $P$ and $S$ waves, the scatter manifests as small values of standard deviation.

\section{Conclusions}
In this paper, we continue the work of~\citet{BosEtAl2018} to investigate the sole mathematical approximation made by~\citet{Backus1962}.
Although it is mathematically possible to achieve a relative error of 100\% for the Backus product approximation, each of the evaluated samples of~\citet{Bormann2012} result in $g>0$\,.
However, within PREM, the first few tens of kilometres do not reflect the actual Earth structure.
Thus, we considered the case study of~\citet{ZhouKaderali2006}, which pertained to a shallow-acquisition region, and concluded that $g>0$ as well.
Thus, in the context of seismology, at both regional and finer scales, potential issues are unlikely to occur as $\overline{g}=0$ requires negative values of $g$\,.

The Backus average might lead to issues when using a less-idealized model or when considering materials with, say, $\nu\approx0$ or $\nu<0$\,.
For the former, this is akin to assuming a symmetry class other than isotropy might result in different formulations of $g$ for which the average is near zero.
For the latter, such materials correspond to the so-called auxetic materials, which may be synthetic or naturally occurring.
Both considerations are addressed by~\citet{Adamus2020}, wherein he derives expressions for $g$ in isotropic, cubic, TI, and tetragonal symmetry classes as well as simulates wave propagation in layered and equivalent media with $g\approx0$\,.
Notably, both considerations are shown, therein, to be much less critical than the requirement that the stack of thin layers must be smaller than the seismic wavelength.
Should this averaging-length requirement not be met, the Backus average is said to be error-laden~\citep[e.g.,][]{SamsWilliamson1994}.
Depending on the purpose of averaging the length, some authors recommend the length be less than or equal to one-third of the seismic dominant wavelength~\citep{LinerFei2006} or that the individual layer thicknesses must be at least ten times smaller than the seismic wavelength~\citep[Section~4.13]{MavkoEtAl2009}.
\section*{Acknowledgments}
We wish to acknowledge discussions with Michael A. Slawinski, as well as the graphic support of Elena Patarini.
This research was performed in the context of The Geomechanics Project supported by Husky Energy. 
Also, this research was partially supported by the Natural Sciences and Engineering Research Council of Canada, grant 238416-2013.
\bibliographystyle{apa}
\bibliography{KKS.bib}
\end{document}